\documentclass{jfm}
\usepackage{graphicx}
\usepackage{epstopdf, epsfig}
\usepackage{blindtext}
\usepackage{longtable}
\usepackage{amsmath}
\usepackage{mathtools}
\usepackage[colorlinks,citecolor = blue, linkcolor=red,hyperindex,CJKbookmarks]{hyperref}

\shorttitle{Subcritical behaviour in double diffusive convection within the diffusive regime}
\shortauthor{K. L. Chong \textit{et al.}}
\title{Subcritical behaviour in double diffusive convection within the diffusive regime}
\author{Kai Leong Chong\aff{1}\corresp{\email{k.l.chong@utwente.nl}},
      Rui Yang\aff{1},
      Yantao Yang\aff{2},
      Roberto Verzicco\aff{1,3,4}
      \and Detlef Lohse\aff{1,5}
 }
\affiliation{
\aff{1}Physics of Fluids Group, Max Planck Center for Complex Fluid Dynamics, MESA+ Institute and J.M.Burgers Center for Fluid Dynamics, University of Twente, P.O. Box 217, 7500 AE Enschede, The Netherlands
\aff{2}SKLTCS and Department of Mechanics and Engineering Science, BIC-ESAT, College of Engineering, and Institute of Ocean Research, Peking University, Beijing 100871, China
\aff{3}Dipartimento di Ingegneria Industriale, University of Rome `Tor Vergata', Via del Politecnico 1, Roma 00133, Italy
\aff{4}Gran Sasso Science Institute - Viale F. Crispi, 7 67100 L'Aquila, Italy
\aff{5}Max Planck Institute for Dynamics and Self-Organisation, 37077 G\"ottingen, Germany}
\begin{document}

\maketitle

\begin{abstract}
We conduct two- and three-dimensional simulations for double diffusive convection in the diffusive regime, where the fluid flow is driven by a destabilizing  temperature gradient and stabilized by a stably stratified salinity gradient. We study how the heat flux, Reynolds number, and flow structures change with the density ratio $\Lambda$, which is the ratio of the buoyancy force induced by the salinity gradient to that by the  temperature gradient. When $\Lambda$ increases from zero, the flow first behaves similarly as in pure Rayleigh-B\'enard (RB) convection, both with respect to flow structure and to heat transport. The linear stability analysis of Baines \& Gill (J. Fluid Mech., vol. 37, 1969, pp. 289--306) had estimated the critical density ratio $\Lambda_c$, above which the flow becomes stable. However, here we show that by using a large-scale circulation as initial condition (rather than the linear profiles assumed in the linear stability analysis), DDC in the diffusive regime can exhibit \emph{subcritical} behaviour when $\Lambda > \Lambda_c$, i.e., coexistence of states at the same control parameters. Even though the density ratio becomes thousands times that of the critical value $\Lambda_c$, there is still convection with strongly enhanced heat transfer properties compared to the pure conduction case. We reveal the corresponding flow structures and find an unstably-stratified region sandwiched between two stably-stratified layers. Our results demonstrate the importance of the initial condition for DDC in the diffusive regime, especially in the situation of a large density ratio, which occurs in high-latitude ocean regions.

\end{abstract}

\begin{keywords}
\end{keywords}

\section{Introduction}
In double diffusive convection (DDC) the fluid density is determined by two scalar components, which usually have different molecular diffusivities \citep{Turner1974,huppert1981double,turner1985multicomponent,Schmitt1994,Radko2013,Garaud2018a}. DDC is ubiquitous in the geophysical systems, such as thermally active lakes \citep{Sommer2013,Sommer2014}, inland seas \citep{Bouffard2019}, and the atmosphere \citep{DoswellIii2008}. In particular, it has strong relevance to the ocean, where the density of the seawater depends on both salinity and temperature \citep{Turner1974,Schmitt1994}. Instability can arise when the destabilizing gradient is strong enough, despite the fact that it is partially counteracted by the second scalar with stable stratification.

\begin{figure}
\centering
\centerline{\includegraphics[width=1.0\textwidth]{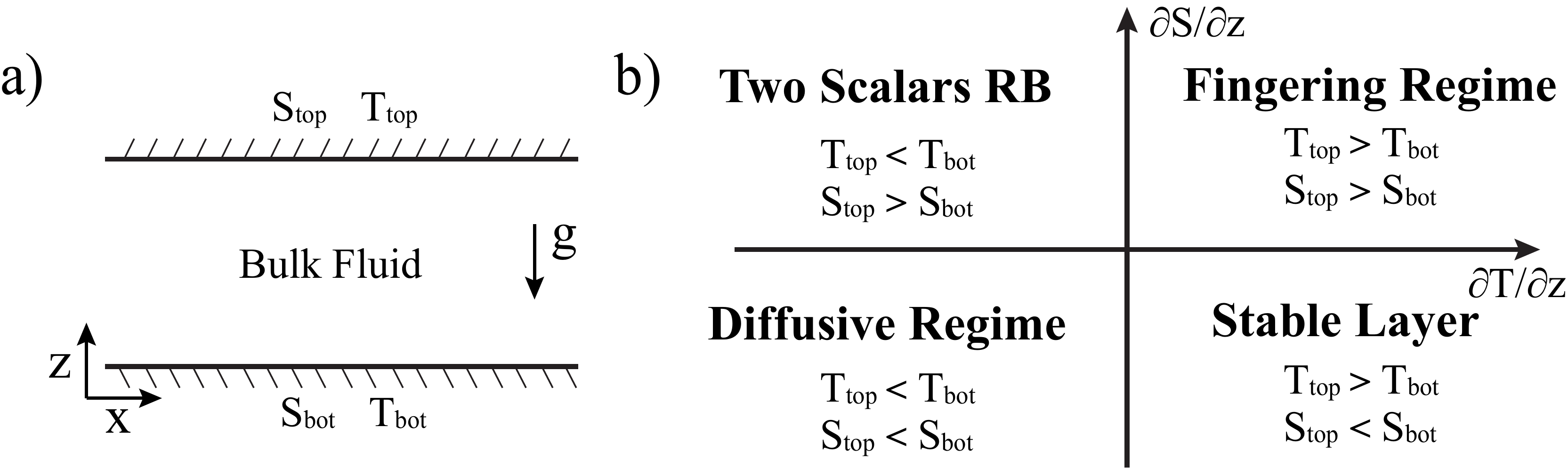}}
\caption{(a) Schematic configuration of the double diffusive convection. (b) Definition of the four different regimes in the double diffusive convection.}
\label{fig:setup}
\end{figure}

In different parts of the oceans, double diffusive convection can belong to different regimes, most importantly the fingering regime and the diffusive regime (see figure \ref {fig:setup}(a,b) for the definition of the regimes). The fingering regime occurs in the tropic and the sub-tropic ocean since the heating of the sunlight over the upper ocean layer and the resulting water evaporation leads to hot salty water, stratified over cold fresh water \citep{Schmitt2005}. It results in the formation of sinking ``salt fingers" (also known as ``salt-fountains") in this regime \citep{Stern1960}. The opposite situation has been found in the high-latitude oceans such as the Canadian Basin \citep{Timmermans2008}. In those regions of the ocean, due to rain and cold surface temperature, the temperature becomes the driving source which opposes the stably-stratified salinity field. The diffusive regime in those regions is thought to maintain the staircase flow structures existing in the ocean, which is a series of well-mixed horizontal layers with sharp interfaces in between \citep{Radko2013,radko2016thermohaline,brown2019initiation}.

One of the key issues of DDC is to understand how salinity flux and heat flux depend on the control parameters. The pioneering work by \cite{turner1965coupled} used the turbulent heat flux scaling from Rayleigh-B\'enard convection to theoretically predict the fluxes in DDC. It gives the relationship between the heat flux $Q_T$ and the temperature difference $\Delta_T$ between the layers, resulting in the scaling relation $Q_T\sim \Delta_T^{4/3}$. This ``4/3 scaling" has been widely used by the oceanographers for estimating the fluxes in the ocean. It can also be used to estimate the salinity flux $Q_S$ in the fingering regime \citep{turner1967salt,schmitt1979flux,mcdougall1984flux,taylor1989laboratory}, implying $Q_S\sim \Delta_S^{4/3}$ where $\Delta_S$ is the salinity difference across the finger layer. However, it is known that there are no single scaling exponents in Rayleigh-B\'enard convection \citep{ahlers2009heat,lohse2010small,chilla2012new}. To obtain a more complete picture of the parameter dependences of the heat \& salinity fluxes, \cite{yang2015salinity,Yang2018} applied the Grossmann-Lohse theory \citep{grossmann2000scaling,grossmann2001thermal,stevens2013unifying}, which is a unified view of the heat fluxes depending on the control parameters, to the double diffusive convection, leading to results in good agreement with their numerical simulations. 

Besides on the dependence on the temperature and salinity difference, $\Delta_T$ and $\Delta_S$, recent simulations have also focused on how the fluxes depend on the density ratio $\Lambda$, namely the ratio of the density change induced by one scalar component to another. Interestingly, for DDC in the fingering regime, \cite{Yang2015} have found that the salinity flux does not decrease monotonically with increasing strength $\Delta_T$ of the stabilization through the temperature gradient. Instead, they find an intermediate regime in which the salinity flux increases with the degree of stabilization through the temperature field, despite that the flow speed is suppressed by the additional stabilizing temperature field. Later, \cite{chong2017confined} proposed a unified view on the flux enhancement in such convective flow with an additional stabilizing force. They report that the enhancement is produced by the increased plume coherence, similarly as in Rayleigh-B\'enard (RB) convection under increasing geometrical confinement \citep{huang2013confinement,chong2015condensation}, rotating RB through onsetting rotation \citep{rossby1969study,zhong2009prandtl,stevens2009transitions} and quasistatic magnetoconvection with increasing magnetic field \citep{lim2019quasistatic}.

In this paper, we focus on the diffusive regime and study how the heat flux, the Reynolds number and the flow structures depend on the density ratio. \cite{Baines1969} performed a linear stability analysis for a basic state with initial linear temperature and salinity gradients, which predicted a critical density ratio \citep{Mirouh2012} $\Lambda_c=(Pr_T+1)/(Pr_T+Le^{-1})$, above which the fluid becomes absolutely quiescent. Here the temperature Prandtl number $Pr_T$ is the ratio of the kinematic viscosity $\nu$ to the thermal diffusivity $\kappa_T$ and the Lewis number $Le$ is the ratio of the thermal diffusivity to the salinity diffusivity defined as $Le=\kappa_T/\kappa_S$. In the ocean, the diffusivity ratio $Le \approx 100$ and $Pr_T \approx 7$, and thus the critical density ratio $\Lambda_c \approx 1.14$. However, in many parts of the ocean, the diffusive regime usually appears in the form of diffusive layering with a density ratio larger than the critical value \citep{Kelley2003}. This considerable contrast  has motivated us to study the influence of the finite-amplitude initial condition on the diffusive regime.

The paper is organized as follows. After a description of the governing equations and the control parameters in Section 2, the details of the numerical method and the numerical set-up are provided in Section 3. Then the results for the flow responses, which are the heat flux and the flow field velocity, are discussed in Section 4. We also demonstrate the existence of subcritical behaviour for DDC in the diffusive regime. To understand the observed behaviour for the heat flux, we further examine the temperature and salinity profiles for different cases in Section 5. Next, we examine the flow structures in Section 6. To understand the existence of the newly found sandwiched convection (an unstably stratified layer bounded between two stably stratified layers) in the diffusive regime, we examine the density profiles in Section 7. The paper ends with conclusions and an outlook (Section 8).

\section{Governing equations and parameters}\label{sec:eq}
We consider the fluid flow between two parallel plates, which are maintained at fixed temperature and salinity difference, $\Delta_T$ and $\Delta_S$, respectively. As these differences are small, the Oberbeck-Boussinesq (OB) approximation can been employed here, such that the fluid density depends linearly on temperature $\tilde{T}$ and salinity $\tilde{S}$,
\begin{equation}
 \tilde{\rho} ( \tilde{T} , \tilde{S} ) = \tilde{\rho} _ { 0 } \left[ 1 - \beta _ {T} \left( \tilde{T} - \tilde{T} _ { 0 } \right) + \beta _ {S} \left( \tilde{S} - \tilde{S} _ { 0 } \right) \right],
\end{equation}
where $\tilde{\rho}_0$, $\tilde{T}_0$, $\tilde{S}_0$ represent the reference density, temperature and salinity, respectively. $\beta_T$ and $\beta_S$ are the thermal and solutal expansion coefficients. With the OB approximation, the governing equations (in dimensionless form) for an incompressible double diffusive convection problem are given by
\begin{equation} \label{eq:mom}
  { \partial _ { t } u _ { i } + u _ { j } \partial _ { j } u _ { i } = - \partial _ { i } p + \sqrt { \frac { P r _ { T } } { R a _ { T } } } \partial _ { j } \partial _ { j } u _ { i } + \left( T - \Lambda S \right) \delta _ { i z } },
\end{equation}
\begin{equation}
 { \partial _ { t } T + u _ { i } \partial _ { i } T = \frac { 1 } { \sqrt { R a _ { T } P r _ { T } }} \partial _ { j } ^ { 2 } T },
\end{equation}
\begin{equation}\label{eq:salinity}
 { \partial _ { t } S + u _ { i } \partial _ { i } S =\frac { 1 } { Le \sqrt { R a _ { T } P r _ { T } } } \partial _ { j } ^ { 2 } S },
\end{equation}
\begin{equation}\label{eq:incom}
\partial _ { i } u _ { i } = 0.
\end{equation}
Here, $u_i$ are the velocity components, $p$ the kinematic pressure, $T$ the temperature and $S$ the salinity. The governing equations were made dimensionless using the domain height $L$, the free-fall velocity $U = \sqrt { g \beta _ {T } | \Delta_T | L }$, and the temperature (or salinity) difference between the top and the bottom plates $\Delta_T$ (or $\Delta_S$). The Kronecker delta $\delta_{iz}$ denotes that the gravitational acceleration $g$ is along the vertical direction.

As seen in the equations, there are four independent dimensionless control parameters, which can be chosen as the Rayleigh and Prandtl number for temperature, the density ratio and the Lewis number:
\begin{align}
  R a _ { T } = \frac { g \beta _ { T } L ^ { 3 } \Delta _ { T } } { \kappa _ { T } \nu } &, {\quad  Pr } _ { T } = \frac { \nu } { \kappa _ { T } }, \\
 \Lambda = \left( \beta _ { S } \Delta _ { S } \right) / \left( \beta _ { T } \Delta _ { T } \right) = R a _ { S } R a _ { T } ^ { - 1 }L e  ^ { - 1 }&, \quad  L e = \kappa _ { T } / \kappa _ { S } =  { Pr } _ { S } P r _ { T } ^ { - 1 },
\end{align}
where $R a _ { S } = { g \beta _ { S } L ^ { 3 } \Delta _ { S } } /{ (\kappa _ { S } \nu) }$ is the salinity Rayleigh number and $ { Pr } _ { S } = { \nu } /{ \kappa _ { S } }$ is the Prandtl number for salinity. $\nu$, $\kappa_T$ and $\kappa_S$ are the kinematic viscosity, the thermal diffusivity, and the solutal diffusivity, respectively.

\begin{figure}
\centering
\centerline{\includegraphics[width=0.7\textwidth]{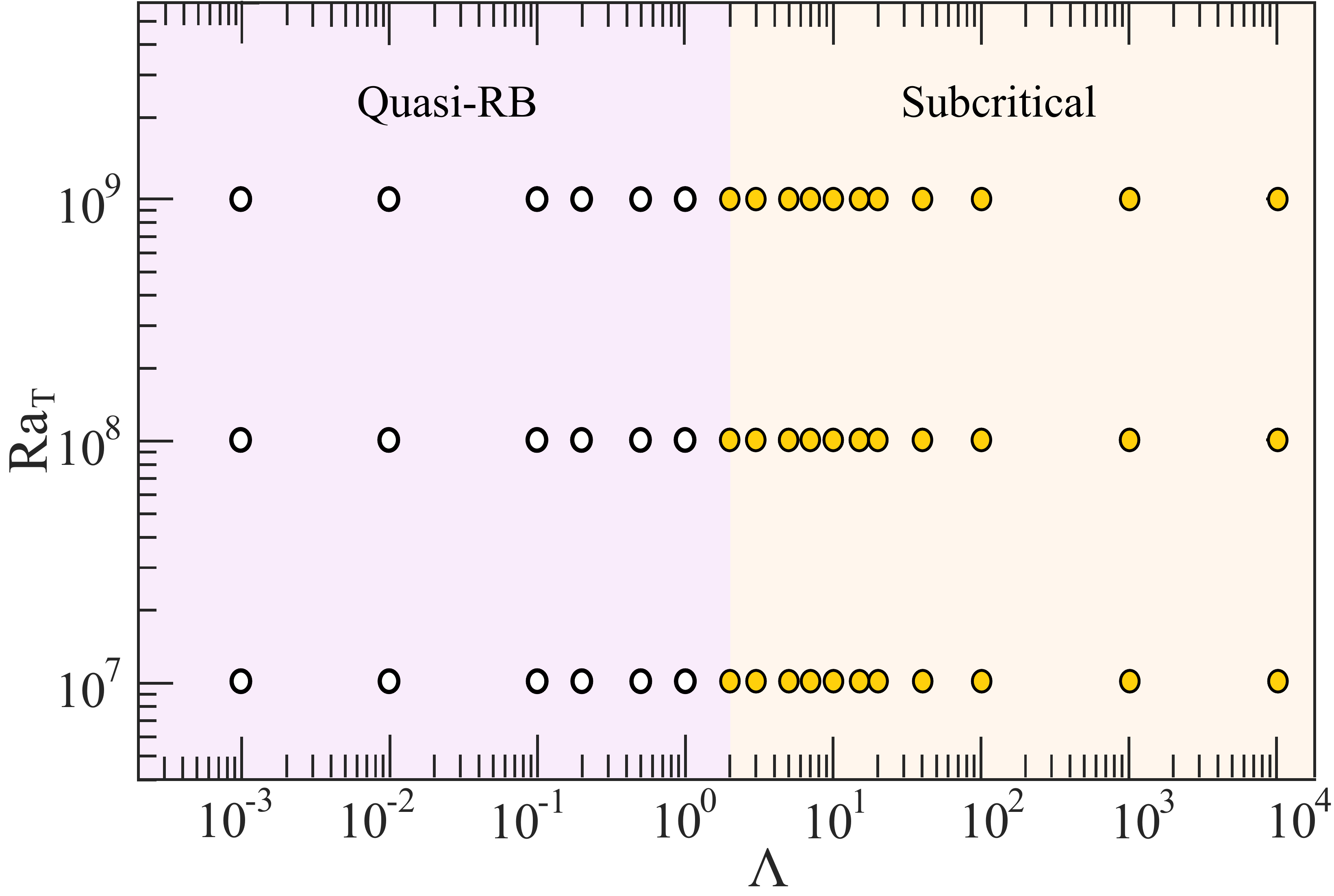}}
\caption{Explored parameter space. White circles denote cases with linear scalar profiles given by equations (\ref{eq:tprofile}) and (\ref{eq:sprofile}) as the initial profiles. Orange circles denote cases conducted with two different types of initial condition, which are the linear scalar profiles and the LSC solution obtained from the pure Rayleigh-B\'enard case. The background colors indicate whether the system behaves in a quasi-RB way or whether it is subcritical, where the boundary $\Lambda\simeq2$ is deduced by the linear stability analysis in \cite{Baines1969}.}
\label{fig:phase}
\end{figure}

The key question in DDC is: How does the system respond to the given control parameters? A particularly important response is the heat flux, or, expressed in non-dimensional form, the temperature Nusselt number $Nu_T$. It can be directly evaluated by
\begin{equation}
Nu_T( z ) = \sqrt {  { Ra_{T}Pr_{T} } } \left\langle u _ { z } T \right\rangle _ {A, t } -\langle \partial_z T\rangle_{A,t},
\end{equation}
where $\langle...\rangle_{A,t}$ denotes the area average at height $z$ and over time. In the statistically stationary case, the z-dependence of $Nu_T$ vanishes. We compute $Nu_T$ at the top and the bottom plates. Another important response parameter is the Reynolds number
\begin{equation}
Re = \frac { \sqrt{ \left \langle \textbf{u}^2  \right \rangle} L } { \nu },
\end{equation}
which characterizes the turbulent intensity of the flow.

\section{Numerical method and set-up}\label{sec:num}
Equations (\ref{eq:mom})--(\ref{eq:incom}) are solved by the second-order finite difference scheme in space and the fractional-step third-order Runge-Kutta scheme combined with the Crank-Nicolson scheme for the implicit terms in time \citep{verzicco1996finite,van2015pencil}. We apply periodic boundary conditions in the horizontal direction, and thus uniform grid spacing has been adopted in this direction. The top and bottom walls are no-slip and with fixed temperature and salinity. A stretched grid is used to resolve the fine structures near the boundaries \citep{shishkina2010boundary}. We also ensure that the grid resolution is fine enough to resolve the Kolmogorov and the Batchelor length scales in the bulk. Details of the resolution for the various cases can be found in appendix A.

Note that with $Le=100$ employed here, the salinity diffuses a hundred times slower than the temperature. Therefore, the resolution for the salinity is much more demanding than that for the temperature. To solve the double diffusive convection in a more efficient way, a multiple-resolution strategy has been employed in which the salinity advection-diffusion equation (\ref{eq:salinity}) is solved on a fine enough grid while the other equations are solved on a base grid. For the details of this method, we refer to \cite{Ostilla-Monico2015}.

We have simulated the two-dimensional cases with Prandtl number for temperature and salinity fixed at $P r_{T}=1$ and $P r_{S}=100$, which corresponds to $Le=100$. Even if $Pr_{S}$ is not as high as in reality, it is two orders of magnitude larger than $Pr_{T}$ and this is enough to capture the strongly different diffusivity effects. Three sets of the temperature Rayleigh number were studied, namely $R a_{T}=10^7, 10^8$, and $10^9$. The salinity Rayleigh numbers $Ra_{S}$ were varied such that the density ratio $\Lambda$ spans between $10^{-3}$ to $10^{4}$. For cases with $\Lambda\geq2$, two different initial conditions were employed. The first one is the linear initial condition, given by
\begin{equation}\label{eq:tprofile}
T(z) = T_{bot} + z(T_{top}-T_{bot}),
\end{equation}
\begin{equation}\label{eq:sprofile}
S(z)= S_{bot} + z(S_{top}-S_{bot}),
\end{equation}
where $z$ is the vertical coordinate ranging from 0 to 1. This initial condition is commonly used in the linear stability analysis. For the second initial condition, we impose the large-scale circulation(LSC) from the pure Rayleigh-B{\'{e}}nard case (i.e. $\Lambda=0$) at the corresponding $Ra_T$. For each case, sufficient statistics was collected after the system had reached the statistical steady state. The explored phase space with the different initial conditions is shown in figure \ref{fig:phase} and the detailed parameters are tabulated in Table \ref{tab:sim}.

In this paper, in most cases, we restrict us to the 2D case of DDC in order to keep the required CPU time in the multi-dimensional parameter space feasible. We do realize that certain dynamical aspects in 2D and 3D convections are different; however, the systematic comparative study of \cite{van2013comparison} has shown that in particular for Prandtl numbers larger than $1$, the 2D simulations give a very good representation on what is going on in 3D. Indeed, for the case where we compare 2D and 3D DDC, we find good qualitative agreement. Here, all results presented are for 2D simulation, if not otherwise explicitly stated.

\section{Fluxes and Reynolds number}
\begin{figure}
\centering
\centerline{\includegraphics[width=0.65\textwidth]{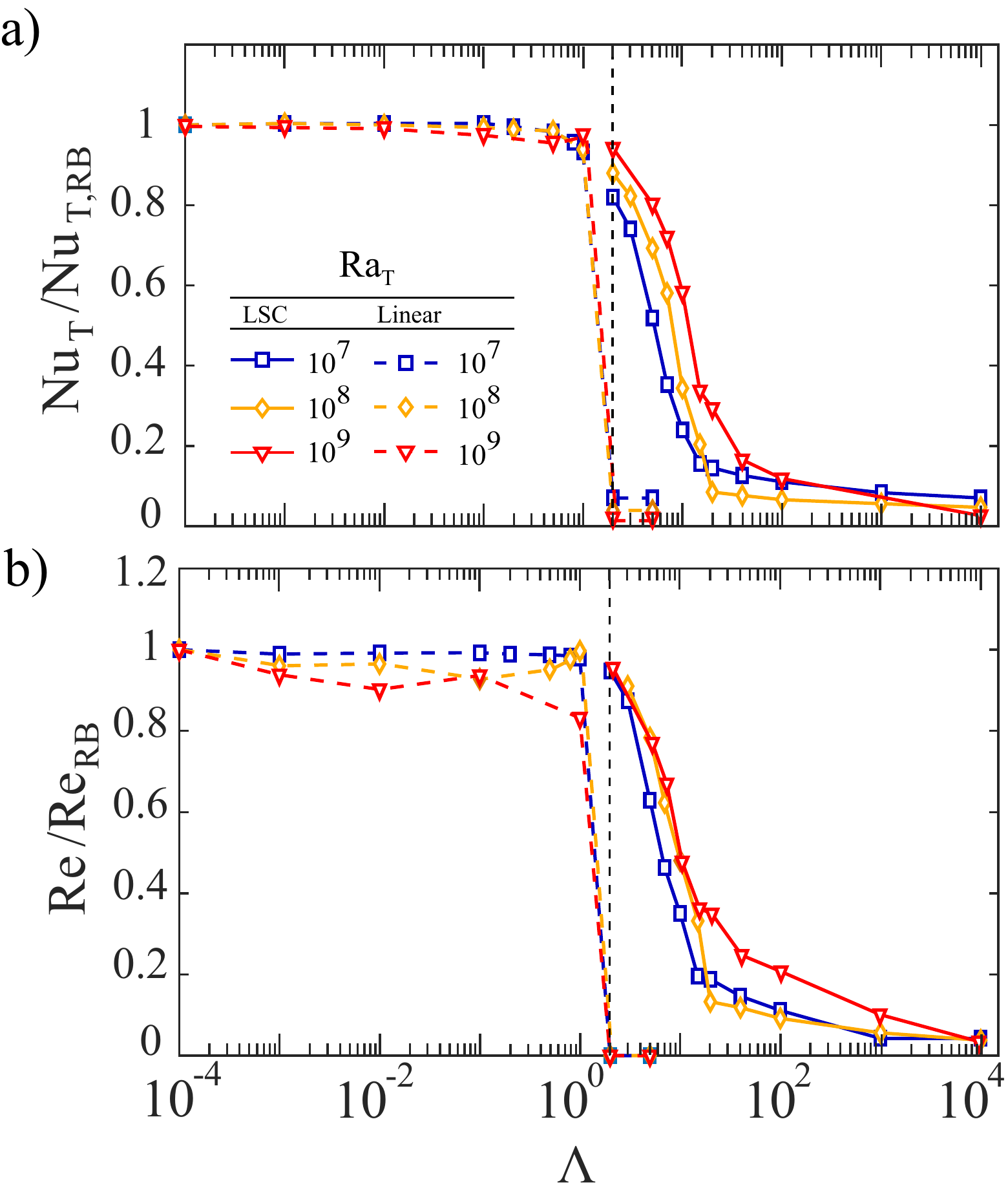}}
\caption{(a) Normalized temperature Nusselt number $Nu_T/Nu_{T,RB}$ and (b) normalized Reynolds number $Re/Re_{RB}$ versus the density ratio $\Lambda$ for $Ra_T = 10^7, 10^8$ and $10^9$ with the linear and the LSC initial conditions. $Nu_{T,RB}$ and $Re_{RB}$ are the Nusselt number and the Reynolds number obtained from the pure Rayleigh-B\'enard case at the respective $Ra_T$. The vertical dashed line indicates the critical density ratio $\Lambda_c$($\simeq 2$) obtained from the linear stability analysis.}
\label{fig:NuRe}
\end{figure}

Figure \ref {fig:NuRe} shows how the heat fluxes and turbulent intensities depend on the density ratio $\Lambda$. The values of  $Nu_T$ and $Re$ are normalized by the corresponding values of the Rayleigh-B{\'{e}}nard case, which are $Nu_{T,RB}$ and $Re_{RB}$. For $\Lambda\leq1$, the stabilizating effect of the salinity gradient on the thermally driven turbulence is still weak. We observe that the values of $Nu_T$ are not sensitive to the change of $\Lambda$ in this regime, which stays close to $Nu_T^{RB}$.

\begin{figure}
\centering
\centerline{\includegraphics[width=0.6\textwidth]{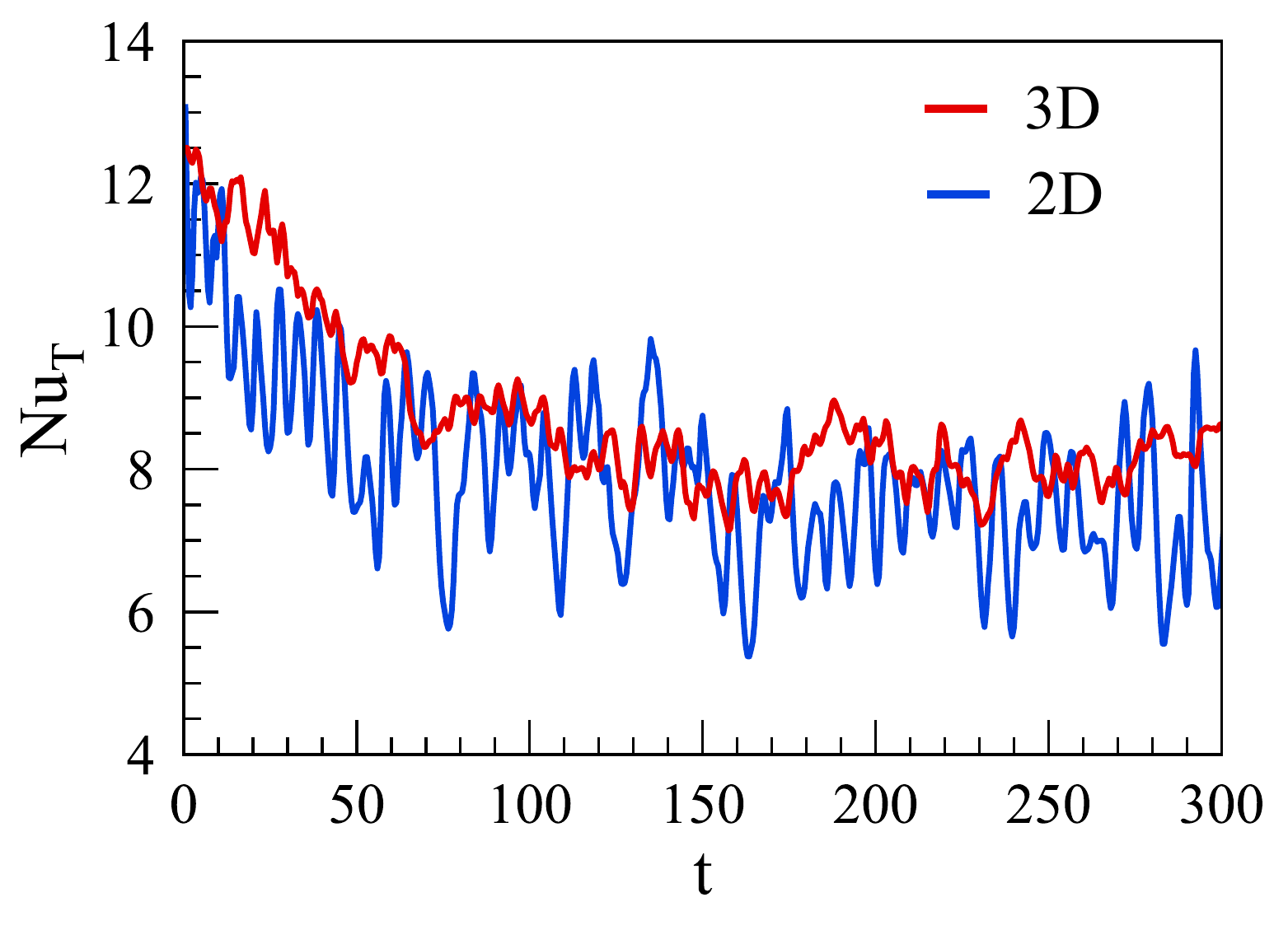}}
\caption{Time series for the temperature Nusselt number $Nu_{T}$ (average of $Nu_{T,top}$ and $Nu_{T,bot}$) for a 2D (blue) and 3D (red) case both at $Ra_T = 10^7$ for $\Lambda=5$ at $Le=100$. The solution obtained from the $\Lambda = 0$ case is used as the initial condition. One sees that the laminar case with $Nu_T=1$, which for $\Lambda>\Lambda_c\simeq2$ one would expect from stability analysis of the linear temperature profile, is not approached and the heat flux remains at a statistically stable value of $Nu_T\simeq8\gg1$ for both 2D and 3D.}
\label{fig:prove}
\end{figure}
\begin{figure}
\centering
\centerline{\includegraphics[width=1.0\textwidth]{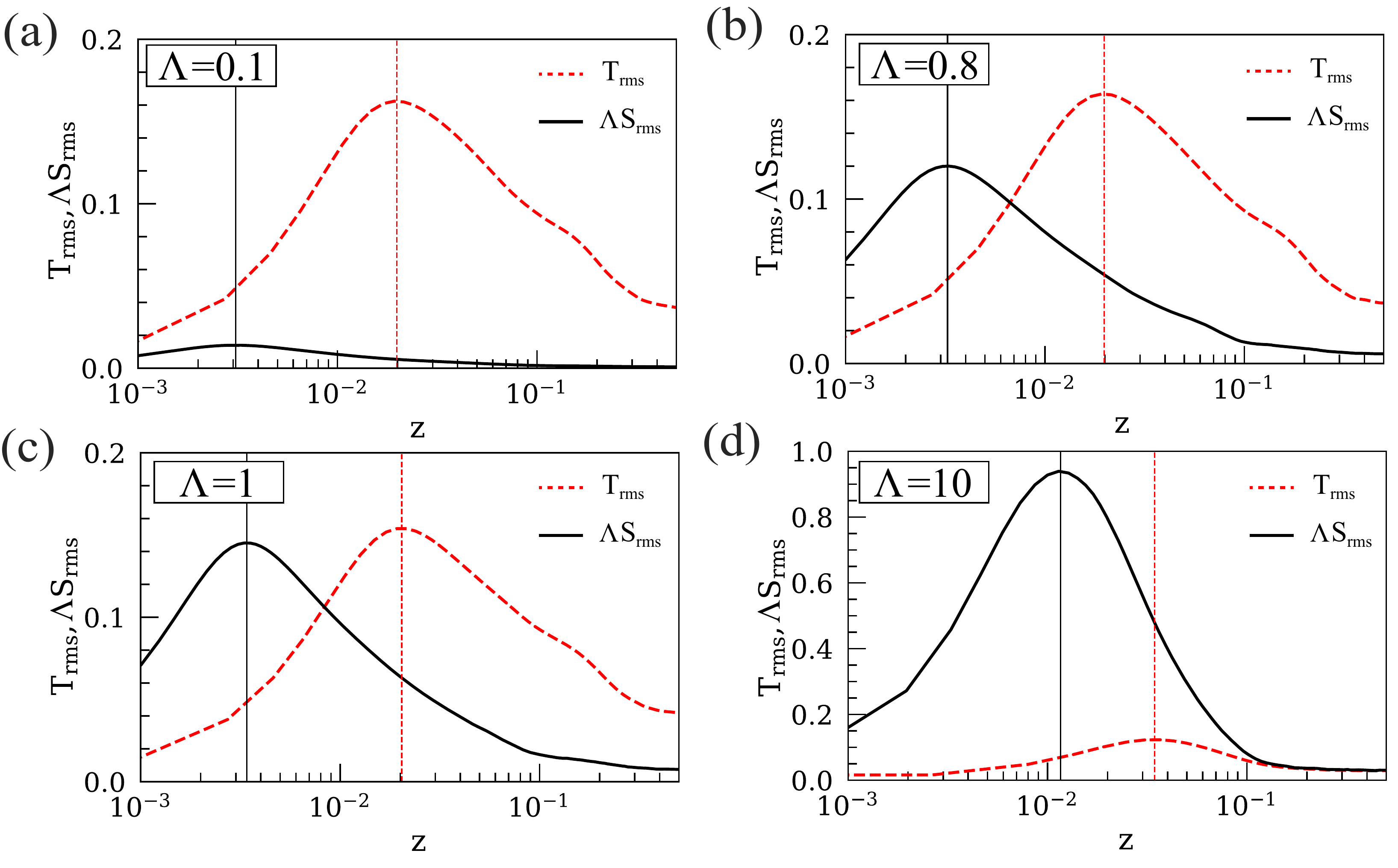}}
\caption{Root-mean-square (rms) profiles of the temperature $T_{rms}$ and the rescaled salinity $\Lambda S_{rms}$ versus height $z$ for (a) $\Lambda = 0.1$; (b) $\Lambda = 0.8$; (c) $\Lambda = 1$ and (d) $\Lambda = 10$ at $Ra_T = 10^8$. Vertical black solid lines indicate the edge of the salinity boundary layer while vertical red dashed lines indicate the edge of the thermal boundary layer, both defined by the maximum of the fluctuations.}
\label{fig:profilerms}
\end{figure}

When $\Lambda$ increases above $2$, we observe subcritical behaviour, i.e., the coexistence of states at the same control parameters. The value of $Nu_T$ now depends on the initial conditions, displaying a lower and an upper branch. For the lower branch, i.e. the cases with linear initial profiles given by the expressions (\ref{eq:tprofile}) and (\ref{eq:sprofile}), our results are in good agreement with the linear stability analysis by \cite{Baines1969}. According to their analysis, for $Pr_T=1$ and $Pr_S=100$, there exists a critical density ratio $\Lambda_c$($\simeq 2$ in this case, and holds for both 2D and 3D) above which the double diffusive convection with the linear initial profile is stable. Our results also suggest that $Nu_T$ sharply declines to $1$ (pure conduction) when $\Lambda$ is above $2$. However, for the upper branch, i.e. with the LSC as initial condition, the $Nu_T$ does not decrease to $1$ immediately after $\Lambda$ reaching $2$. The instability can still persist even though the stabilization by the salinity becomes much stronger than the driving by the temperature field ($\Lambda$ up to $10^4$ in this study). 

How robust is our result, in particular, is the subcritical behaviour also present in 3D? To find out, we performed 3D numerical simulations at $Ra_T=10^7$, $Le=100$ and $\Lambda=5$. Figure \ref {fig:prove} shows that with the large-scale circulation as the initial condition (the 3D RB case), $Nu_T$ does not decline to $1$ even when $\Lambda>\Lambda_c$, signaling the same subcritical behaviour which we have observed in 2D.

\begin{figure}
\centering
\centerline{\includegraphics[width=0.6\textwidth]{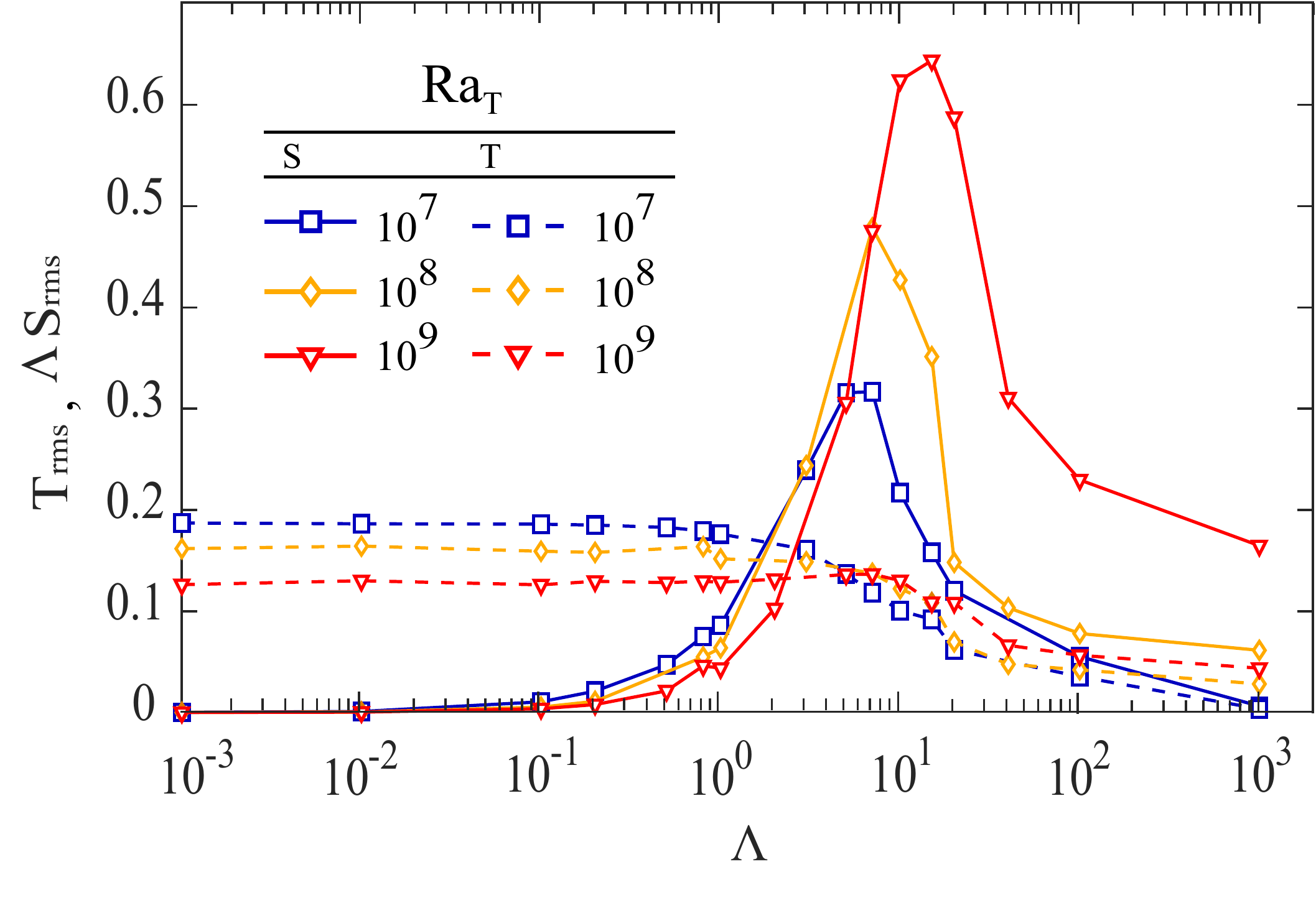}}
\caption{Root-mean-square (rms) values of the temperature (dashed line) and the rescaled salinity (solid line) at the edge of the \emph{thermal} boundary layer, i.e., where the temperature fluctuations are maximal. Note that at that position, in general, the salinity fluctuations are not maximal, as the salinity boundary layer is much thinner than the thermal one, see figure \ref{fig:profilerms}.}
\label{fig:sigma}
\end{figure}

\section{Temperature and salinity fluctuations}
For both, the lower and upper branches, the heat transport and the turbulent intensity decrease for $\Lambda>2$; extremely sharply for the lower branch, and less sharply for the upper one. Two questions arises: (i) Why does the effect of the stabilizing force on global quantities appear all of the sudden? (ii) Why is $Nu_T$ almost unchanged at its value at $\Lambda \simeq 0$ even close to the transition point $\Lambda=1$, although $\Lambda$ is the global measure of the relative contributions of the temperature and the salinity?

In the following, we will explain this decreasing trend of $Nu_T$ and $Re$ by the local competition between salinity and temperature effects. Figure \ref {fig:profilerms} shows how the temperature and the salinity (rescaled by $\Lambda$) root-mean-square values vary with height $z$. When $\Lambda<1$, at the edge of the thermal boundary layer, the rescaled salinity fluctuations are negligibly small. It means that the emitted thermal plumes can hardly experience the stabilization effect from the stably-stratified salinity field. With increasing $\Lambda$, there is increasing stabilization from the salinity. When $\Lambda$ becomes $1$, the peak values of $T_{rms}$ and $\Lambda S_{rms}$ become comparable. However, since the salinity boundary layer is nested inside the thermal one (remember that $Le=\kappa_T/\kappa_S=100$), the emitted thermal plumes do not experience the strongest stabilization from the salinity field, because $\Lambda S_{rms}$ is still three times smaller than $T_{rms}$. Until $\Lambda$ has increased to a much larger value, at which the stabilization dominates over the thermal driving, the global heat transport decreases considerably.

To quantify the relative strength of the thermal driving and the stabilization from the salinity field, we directly measure the value of the salinity fluctuations rescaled by $\Lambda$ at the edge of the thermal boundary layer. Figure \ref {fig:sigma} clearly shows that at low $\Lambda$, the strengths of the rescaled salinity fluctuations $\Lambda S_{rms}$ are negligibly small compared to $T_{rms}$, leading to the almost unchanged $Nu_T$ at small enough $\Lambda$. However, $\Lambda S_{rms}$ increases sharply and overcomes $T_{rms}$ around $\Lambda = 2 $, such that the stabilizing force becomes effective at such $\Lambda$. Indeed, $Nu_T$ and $Re$ both decrease sharply beyond this density ratio. Our results show that these local properties at the thermal boundary layers are closely connected to the global transport behaviours observed in figure \ref {fig:NuRe}.

\begin{figure}
\centering
\centerline{\includegraphics[width=1.0\textwidth]{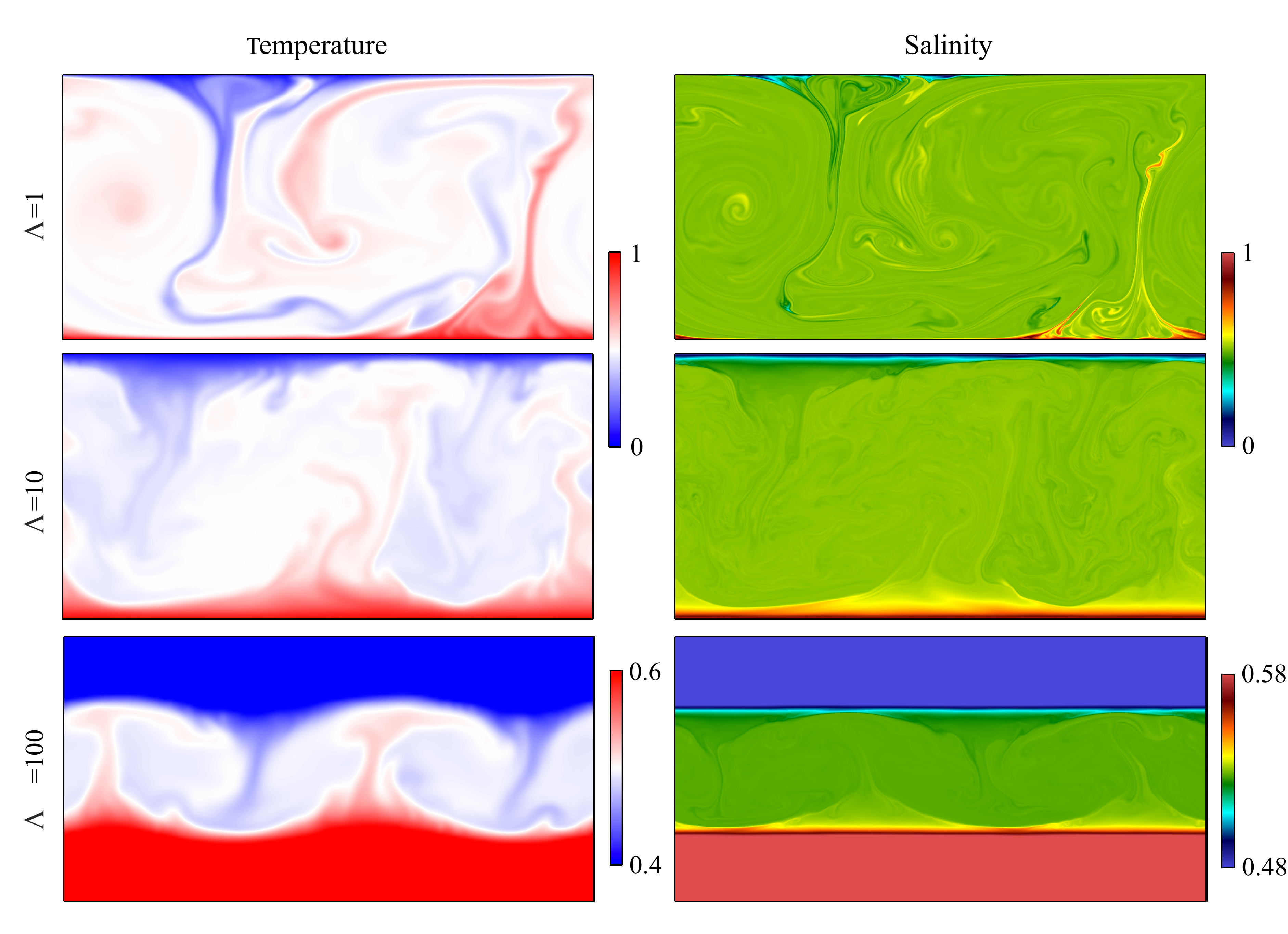}}
\caption{Instantaneous snapshots of salinity (right) and temperature (left) fields at $Ra_T = 10^8$ and $\Lambda = 1, 10$ and $100$. For the temperature field, bluish (reddish) color represents colder (hotter) fluid. For the salinity field, bluish (reddish) color represents fresher (saltier) fluid. For large $\Lambda$, clear layering can be observed.}
\label{fig:morphTS}
\end{figure}

\section{Flow morphologies}
We now examine the corresponding flow morphologies for the state in the upper branch. The temperature and the salinity snapshots are shown in figure \ref {fig:morphTS} at $Ra_T=10^8$ with three different density ratios (from top to bottom: $\Lambda = 1, 10$ and $100$). 

\begin{figure}
\centering
\centerline{\includegraphics[width=1.0\textwidth]{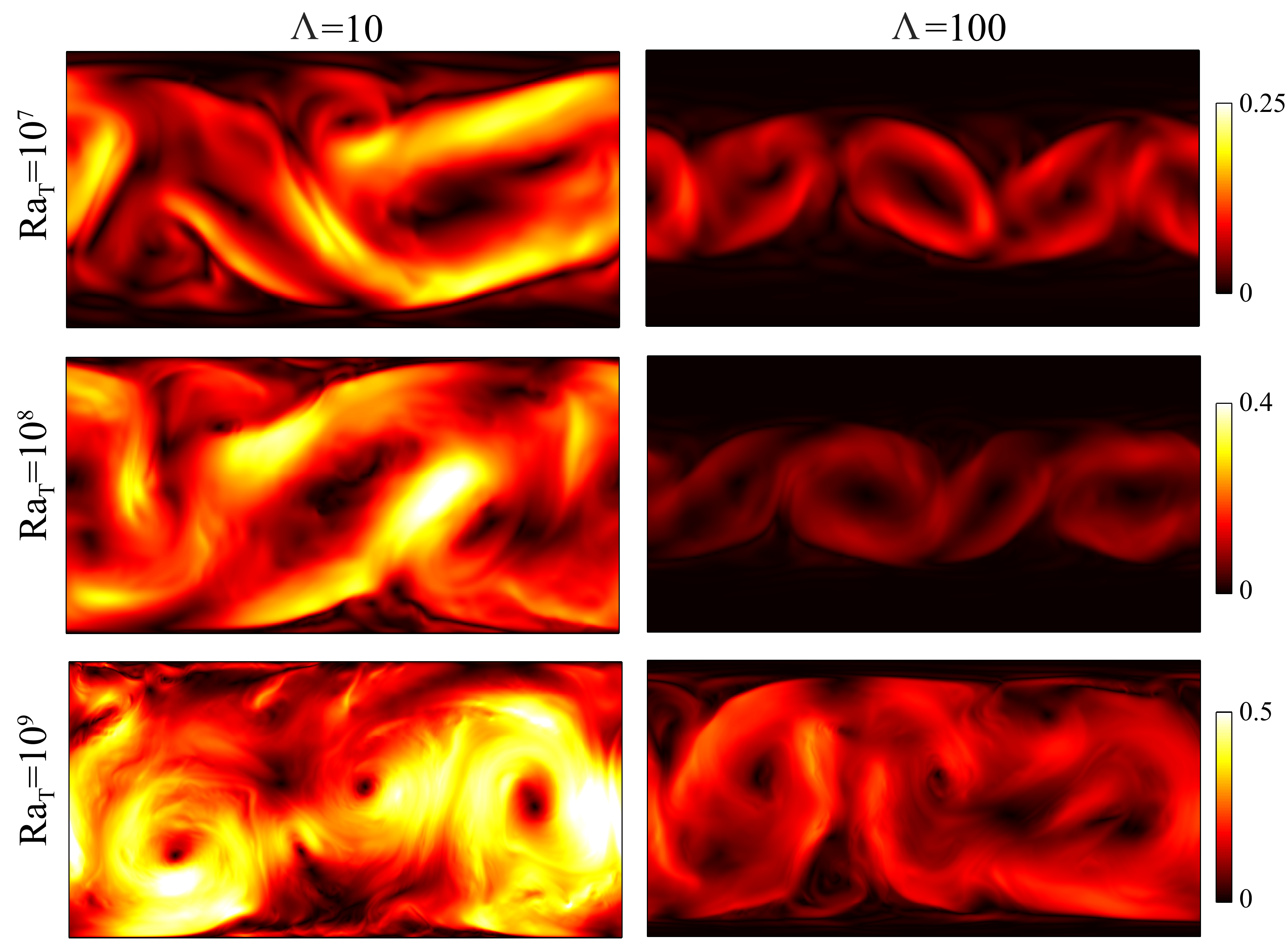}}
\caption{Instantaneous snapshots of the square of the velocity $\textbf{u}^2$ for $Ra_T = 10^7, 10^8, 10^9$ and $\Lambda = 10$ and $100$. It indicates the formation of the stagnation regions (regions with almost zero velocity) adjacent to the top and the bottom plates for large enough $\Lambda$.}
\label{fig:morphV}
\end{figure}

At $\Lambda=1$, the large-scale circulation is the dominating flow structure, similar to the case of pure Rayleigh-B{\'{e}}nard convection. Mushroom-like plumes carry heat from the boundary layer to the bulk up to the opposite plate. These plumes also self-organize themselves into the large-scale circulation (LSC), where the hot plumes traverse upward on one side while the cold ones traverse downward on another side. The footprint of the LSC is also observed in the salinity field. However, the salinity field is passively advected in this case as one can observe that the fresh (less dense) fluid is advected downward by the LSC while the salty (denser) fluid is advected upward by the LSC.

With $\Lambda$ increasing to $10$, which is much larger than the critical value ($\Lambda_c \simeq 2$), the LSC can still persist. It contrasts with the motionless solution in the lower branch. Our results show that the energy can still feed into the LSC even though the strong stabilizing salinity is supposed to damp out the LSC. In this regime, the temperature is still the driving source of the LSC while the salinity field is passively advected by the LSC. For even larger $\Lambda$ ($=100$), it is surprising to observe that the LSC is still active in the largely-narrowed bulk region. With the density ratio increasing further, the region of the bulk will be continuously narrowed until the whole domain becomes stable without any active flow.

Accompanied by the snapshots of the square velocity in figure \ref {fig:morphV}, we can further reveal the formation of stagnation regions (i.e. the region of almost zero velocity) near the top and bottom plates. Take $Ra_T=10^7$ as an example: When the density ratio increases to $100$, the portion of the stagnation zone increases to around half of the domain. At the same time, the strength of the large scale circulation has also been weakened thanks to the increased density ratio. For larger $Ra_T$, the ``dead fluid" zones become less prominent, owing to the more turbulent flow, although they can still be seen for large enough $\Lambda$.

\section{Temperature, salinity and density profiles}
\begin{figure}
\centering
\centerline{\includegraphics[width=1.0\textwidth]{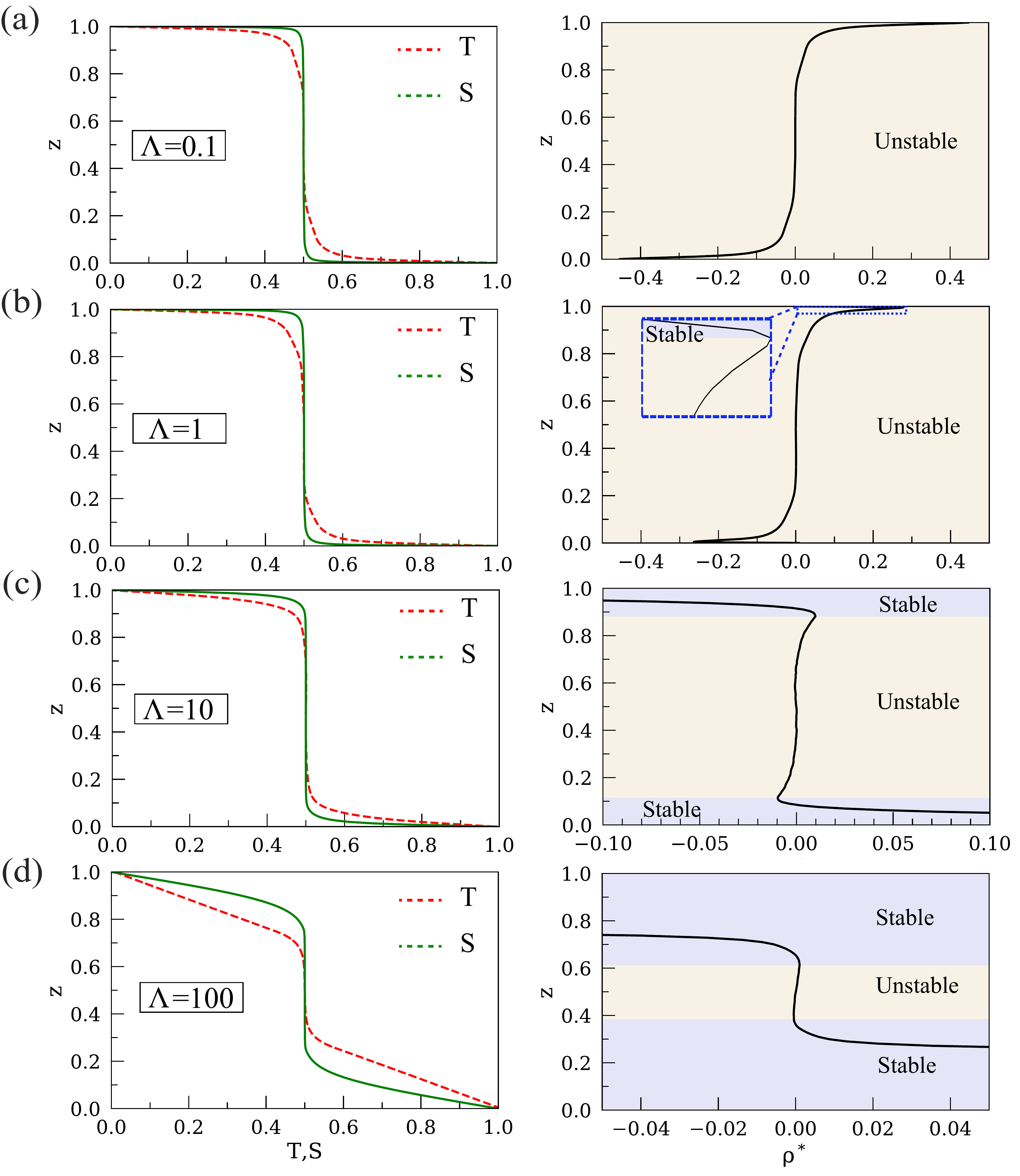}}
\caption{Temperature, salinity and density vertical profiles for $\Lambda = 0.1, 1, 10$ and $100$ at $Ra_T = 10^8$.}
\label{fig:rho}
\end{figure}

To understand why the large-scale circulation can be sustained for $\Lambda$ larger than the critical value $\Lambda_c \simeq 2$ (upper branch of $Nu$) and why stagnation zones form, we examine the density profile as shown in figure \ref{fig:rho}. The expression is given by
\begin{equation}
\rho^*(z)=(\langle T\rangle_{A(z),t}-\langle T\rangle_{A(z=0.5),t}) - \Lambda (\langle S\rangle_{A(z),t}-\langle S\rangle_{A(z=0.5),t}),
\end{equation}
where the symbol $\langle ...\rangle_{A(z=0.5),t}$ denotes the area average at the height $z=0.5$ and over time. For comparison, the density profile solely based on the temperature and the salinity profile are also shown in figure \ref{fig:rho}.

Figure \ref{fig:rho} (a) shows that, for $\Lambda=0.1$, the density profiles just resemble those for the case of pure Rayleigh-B\'enard convection. The density mainly varies within the thin boundary layer whereas there is a density short-cut within the bulk region, i.e. the density has a constant value. The formation of the well-mixed region is attributed to the overturning flow in the bulk, similar to the formation of the thermal short-cut in the bulk of RB convection \citep{ahlers2009heat}. At this small enough $\Lambda$, the stabilizing effect from the salinity field is still negligible, and thus the density field is unstably-stratified over the entire cell height.

When $\Lambda$ increases to $1$, there are two additional stably-stratified layers developed near the top and the bottom plate. They can be clearly seen if one zooms into the region near the plates as shown in figure \ref{fig:rho} (b) for the upper layer. With $\Lambda$ increasing to $10$, the thickness of the stable layers increases progressively. The stably-stratified layers explain why there are dead fluid zones adjacent to the top and the bottom plates because any motion within these layers will be damped out by the stable stratification therein. However, we still find that there is an gravitationally unstably stratified region remaining in the bulk.

One can understand the presence of this unstably-stratified layer by examining the respective temperature and salinity profiles. First, the thermal boundary layer diffuses more rapidly than the salinity boundary layer: As $Le=\kappa_T/\kappa_S=100$, molecular diffusion will always tend to produce a relatively thicker thermal boundary layer than the salinity one. In the region $0.8\leq z \leq 0.9$ and $0.1\leq z \leq 0.2$, the temperature still varies significantly. However, the salinity has already reached the bulk value. Without any salinity gradient in these regions, it leads to the unstable stratification which maintains the LSC. 

The density profiles in figure \ref{fig:rho} demonstrate that for large enough density ratios, the flow structure in the upper branch is fundamentally different from that of RB convection. In contrast to the entirely unstably-stratified flow in RB convection, there exist two additional stable layers adjacent to the plates while there is unstable stratification sandwiched between the two stable layers. We therefore refer to it as the ``sandwiched" structure to distinguish this flow structure from that in the quasi-RB regime.

\begin{figure}
\centering
\centerline{\includegraphics[width=0.6\textwidth]{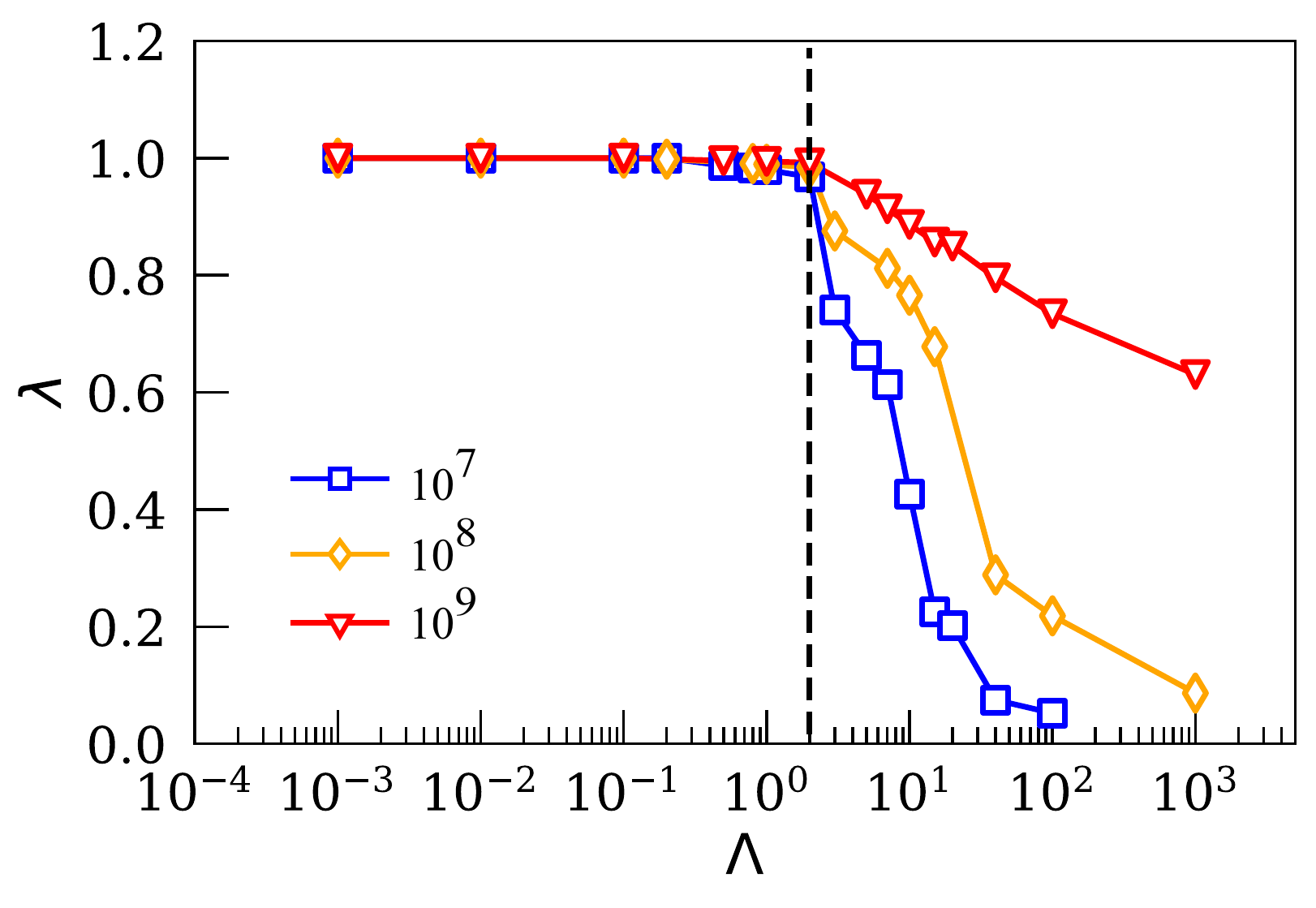}}
\caption{Thickness of unstably-stratified region $\lambda$ versus density ratio $\Lambda$ for $Ra_T = 10^7, 10^8$ and  $10^9$.}
\label{fig:thickness}
\end{figure}

Figure \ref{fig:thickness} quantifies the thickness $\lambda$ of the unstable region, based on the density profile. It is noted that the unstable region starts to decrease rapidly at the critical density ratio ($\Lambda \simeq 2$). But as the Rayleigh number increases, the thickness of the unstable region decreases more slowly. For $Ra_T = 10^9$, the active region is still 60 percent of the whole domain, while for $Ra_T = 10^7$ and $10^8$, the active flow region almost disappears.

\section{Concluding remarks and outlook} \label{sec:conc}
In summary, double diffusive convection in the diffusive regime was studied numerically for a large range of control parameters with $Ra_T=10^7, 10^8, 10^9$, $\Lambda$ between $0$ and $10^4$ and $Le=100$.

With the density ratio $\Lambda$ increasing from zero, the system is first in the quasi-RB regime in which the heat flux is comparable to that in the pure Rayleigh-B\'enard case. When $\Lambda$ becomes larger than $\Lambda_c$, we find a subcritical regime, in which there is coexistence of different states at the very same control parameters, which of these is taken depends on the initial conditions. With linear profiles as the initial conditions (lower branch), the laminar solution with $Nu_T=1$ was obtained. However, with LSC as the initial condition (upper branch), there is still convection with strongly enhanced heat transport properties after reaching the statistical steady state.

Different flow structures were observed in the two regimes. In the quasi-RB regime, LSC is the dominating flow structure. In the subcritical regime, we found that LSC is confined in the largely-narrowed bulk while there are two stagnation regions adjacent to the plates. By examining the density profile, we understood that such a ``confined" flow structure is caused by the unstable stratification sandwiched between the two stably-stratified layers.

Finally, we have demonstrated the subcritical behaviour of DDC in the diffusive regime. Our results imply that the initial conditions play an important role in the flow structures and the response parameters. Although the system is globally stable when the density ratio is larger than the critical value $\lambda_c$, locally in the bulk an unstably-stratified region exists with remaining circulation. So far, we qualitatively showed that our results also hold for one case in 3D. Obviously it is interesting to study more 3D cases in the future to test the multi-dimensional parameter space. Moreover, it is worthwhile to explore the effect of initial conditions with staircase structure, which has strong relevance to the geophysical systems.

\section*{Acknowledgements}
We greatly appreciate valuable discussions with Alexander Blass, Richard Stevens and Qi Wang. We acknowledge the support from an ERC-Advanced Grant under the project number $740479$. K. L. C. acknowledges Croucher Foundation for Croucher Fellowships for Postdoctoral Research. Y. Y. acknowledges the support from the Major Research Plan of National Nature and Science Foundation of China for Turbulent Structures under the Grants 91852107 and 91752202. We also acknowledge PRACE for awarding us access to MareNostrum at the Barcelona Supercomputing Centre (BSC) under PRACE project number 2017174146 and JUWELS at the J\"ulich Supercomputing Centre. This work was also partly carried out on the national e-infrastructure of the SURFsara with the support of SURF Cooperative.

\section*{Declaration of interests}
The authors report no conflict of interest.

\section*{Appendix A. Numerical parameters}
\setlength{\LTcapwidth}{0.9\linewidth}
\begin{longtable}{@{\extracolsep{\fill}}cccccccccc@{}}
\hline\hline
     $Ra_T$   & $\Lambda$  & Initial  & $N_x\times N_z$   &   $n_x\times n_z$ &  $N_T$ & $N_S$ & $t_{avg}$ & $Nu_T$ & \\[3pt]
\hline
\endfirsthead
\hline\hline
     $Ra_T$   & $R_\rho$  & Initial  & $N_x\times N_z$   &   $n_x\times n_z$ &  $N_T$ & $N_S$ & $t_{avg}$ & $Nu_T$ & \\[3pt]
\hline
\endhead
\hline
\endfoot
$10^7$ &	    0 &   Linear   &288$\times$144  & 1$\times$1  &8  &$\backslash$   &  400	& 14.25 \\
       &	0.001 &   Linear   & 288$\times$144  & 6$\times$6  &8  &13  &  400	& 14.29 \\
       &	 0.01 &   Linear   & 288$\times$144  & 6$\times$6  &8  &13  &  400	& 14.29 \\
       &	  0.1 &   Linear   & 288$\times$144  & 6$\times$6  &8  &13  &  350	& 14.30 \\
       &	  0.2 &   Linear   & 288$\times$144  & 6$\times$6  &8  &13  &  350	& 14.24 \\
       &	  0.5 &   Linear   & 288$\times$144  & 6$\times$6  &9  &13  &  350	& 14.00 \\
       &	  0.8 &   Linear   & 288$\times$144  & 6$\times$6  &9  &13  &  350	& 13.63 \\
       &	    1 &   Linear   & 288$\times$144  & 6$\times$6  &9  &14  &  350	& 13.28 \\
       & 	    2 &    LSC     & 288$\times$144  & 6$\times$6  &10 &16  &  400	& 11.69 \\
       & 	    3 &    LSC     & 288$\times$144  & 6$\times$6  &11 &18  &  300	& 10.55 \\
       & 	    5 &    LSC     & 288$\times$144  & 6$\times$6  &12 &32  &  300	&  7.40 \\
       & 	    7 &    LSC     & 288$\times$144  & 6$\times$6  &20 &62  &  300	&  5.04 \\
       & 	   10 &    LSC     & 288$\times$144  & 6$\times$6  &28 &110  &  300	&  3.41 \\
       & 	   15 &    LSC     & 288$\times$144  & 6$\times$6  &39 &173  &  300	&  2.23 \\
       & 	   20 &    LSC     & 240$\times$120  & 4$\times$4  &34 &97  &  300	&  2.07 \\
       & 	   40 &    LSC     & 240$\times$120  & 4$\times$4  &38 &104  &  300	&  1.80 \\
       & 	  100 &    LSC     & 240$\times$120  & 4$\times$4  &42 &104  &  150	&  1.57 \\
       & 	 1000 &    LSC     & 240$\times$120  & 4$\times$4  &52 &112  &  150	&  1.19 \\
       & 	10000 &    LSC     & 240$\times$120  & 4$\times$4  &60 &240  &  150	&  1.00 \\
\hline
$10^8$ &	    0 &   Linear   & 480$\times$240  & 1$\times$1	 &10 &$\backslash$  &  400	& 25.46 \\
       &	0.001 &   Linear   & 480$\times$240  & 8$\times$8  &10 &15  &  400	& 25.56 \\
       &	 0.01 &   Linear   & 480$\times$240  & 8$\times$8  &10 &15  &  400	& 25.48 \\
       &	  0.1 &   Linear   & 480$\times$240  & 8$\times$8  &10 &15  &  400	& 25.31 \\
       &	  0.2 &   Linear   & 480$\times$240  & 8$\times$8  &10 &15  &  400	& 25.19 \\
       &	  0.5 &   Linear   & 480$\times$240  & 8$\times$8  &10 &17  &  400	& 25.09 \\
       &	  0.8 &   Linear   & 480$\times$240  & 8$\times$8  &10 &17  &  400	& 24.80 \\
       &	    1 &   Linear   & 480$\times$240  & 8$\times$8  &11 &18  &  400	& 23.92 \\
       & 	    2 &    LSC     & 480$\times$240  & 8$\times$8  &11 &18  &  250	& 22.80 \\
       & 	    3 &    LSC     & 480$\times$240  & 8$\times$8  &12 &20  &  250	& 20.94 \\
       & 	    5 &    LSC     & 480$\times$240  & 8$\times$8  &14 &26  &  200	& 17.64 \\
       & 	    7 &    LSC     & 480$\times$240  & 8$\times$8  &16 &36  &  200	& 14.79 \\
       & 	   10 &    LSC     & 240$\times$120  & 4$\times$4  &10 &17  &  200	&  8.76 \\
       & 	   15 &    LSC     & 240$\times$120  & 4$\times$4  &16 &36  &  200	&  5.19 \\
       & 	   20 &    LSC     & 240$\times$120  & 4$\times$4  &33 &95  &  200	&  2.16 \\
       & 	   40 &    LSC     & 240$\times$120  & 4$\times$4  &35 &99  &  200	&  1.95 \\
       & 	  100 &    LSC     & 240$\times$120  & 4$\times$4  &40 &108  &  200	&  1.68 \\
       & 	 1000 &    LSC     & 240$\times$120  & 4$\times$4  &45 &108  &  200	&  1.43 \\
       & 	10000 &    LSC     & 240$\times$120  & 4$\times$4  &52 &101  &  200	&  1.20 \\
\hline
$10^9$ &	    0 &   Linear   & 896$\times$448  & 1$\times$1  &15 &$\backslash$  &  200	& 47.68 \\
       &	0.001 &   Linear   & 896$\times$448  & 6$\times$6  &15 &21  &  120	& 47.54 \\
       &	 0.01 &   Linear   & 896$\times$448  & 6$\times$6  &15 &21  &  110	& 47.40 \\
       &	  0.1 &   Linear   & 896$\times$448  & 6$\times$6  &15 &22  &  120	& 46.59 \\
       &	  0.2 &   Linear   & 896$\times$448  & 6$\times$6  &15 &23  &  120	& 46.30 \\
       &	  0.5 &   Linear   & 896$\times$448  & 6$\times$6  &15 &23  &  110	& 45.67 \\
       &	    1 &   Linear   & 896$\times$448  & 6$\times$6  &15 &23  &  130	& 46.55 \\
       & 	    2 &    LSC     & 896$\times$448  & 2$\times$2  &15 &8  &  140	& 44.95 \\
       & 	    5 &    LSC     & 896$\times$448  & 2$\times$2  &18 &9  &  140	& 38.31 \\
       & 	    7 &    LSC     & 896$\times$448  & 2$\times$2  &19 &10  &  140	& 34.38 \\
       & 	   10 &    LSC     & 896$\times$448  & 2$\times$2  &23 &14  &  150	& 27.83 \\
       & 	   15 &    LSC     & 896$\times$448  & 2$\times$2  &34 &30  &  140	& 16.09 \\
       & 	   20 &    LSC     & 896$\times$448  & 2$\times$2  &38 &36  &  140	& 14.00 \\
       & 	   40 &    LSC     & 896$\times$448  & 2$\times$2  &57 &77  &  140	&  7.90 \\
       & 	  100 &    LSC     & 896$\times$448  & 2$\times$2  &70 &95  &  140	&  5.66 \\
       & 	 1000 &    LSC     & 896$\times$448  & 2$\times$2  &119 &101 &  140	&  2.59 \\
       & 	10000 &    LSC     & 896$\times$448  & 2$\times$2  &192 &448 &120	&  1.25 \\
 \hline
  \caption{Simulation parameters and the resulting global convective heat flux. The Prandtl number for temperature and salinity are both kept constant at $Pr_T=1$ and $Pr_S=100$, respectively. The aspect ratio is fixed as 2. The columns from left to right indicate the thermal Rayleigh number $Ra_{T}$, the density ratio $\Lambda$, the type of the initial profiles, the number of grid points in horizontal and vertical direction $N_x \times N_z$, the refinement factor in horizontal and vertical directions $n_x \times n_z$, the number of grid points in the thermal boundary layers ($N_T$), the number of grid points in the salinity boundary layers ($N_S$), the averaging time  $t_{avg}$ in free fall time units, and the thermal Nusselt number $Nu_T$. Two types of the initial profiles were employed (i) The linear profiles (Linear) given in the equations (\ref{eq:tprofile}) and (\ref{eq:sprofile}) and (ii) the large-scale circulation (LSC) condition obtained from the pure Rayleigh-B\'enard case. Note that for the cases with $\Lambda \geq2$, both initial conditions were run, but $Nu_T=1$ is only obtained for the linear cases.}
  \label{tab:sim}
\end{longtable}

\end{document}